# Wave model of forming of the martensite crystal in the heterogeneity medium

V. Chashchina, M. Kashchenko, S. Vikharev

*Physics chair, Ural State Forest Engineering University, 620100, Ekaterinburg, Russia*
*E-mail: mpk46@mail.ru*

In the current work considering the wave model of the crystal growth control, an influence of heterogeneity of the medium on the forming martensite crystal s profile was examined. The considering of the heterogeneity is provided with the help of putting into operations the space-depending effective attenuation of the waves. The description of the heterogeneity was fulfilled in three different ways: exponential, quadratic and inverse-quadratic. It was shown, that dependently of the heterogeneity of the medium various martensite crystal s profile can be implemented, considering the shape of its butts as well. In particular the plate-like and the wedge-like of the crystal shapes are feasible.

The plate-like shape can be considered as a typical one for the martensite crystal s, same for thin twin structure (TS) components. However, both the boarders of the crystal, that starts its growth in the presence of heterogeneity (e.g., in case of interaction with crystallite's edges or any other fresh-formed and forming crystals) and the edges of the TS-components are well-known to have the wedge-like, needle-shaped in profile, shape. With getting of the model description of the elastic twinning and forming of the thermoelastic crystals, which speed is considerably less in comparison with velocity of sound, the wedge-like shape of displacing crystallites can be easily interpreted on the base of the dislocation approach [1]. However, such an interpretation is not acceptable in terms of supersonic velocity of the martensite crystals, specified in the framework of the controlling wave process (CWP) concept [2,3]. Let us consider, according to [2], that in the lattice area, which becomes instable with the CWP influence, nonequilibrium electronic subsystem provides the maintenance of the high level amplitude vibrations and corresponding deformations as well. Thus, with no specification of the microscopic mechanism of generation (enhancement), we suppose the presence of compensation of the directive wave's attenuation in the area, where deformations of propagation $\varepsilon_1 > 0$ and compression $\varepsilon_2 < 0$ overlay. Let us consider some deformation equations to be the basis (for the deformations that are displacing by the waves in the orthogonal $n_1$- и $n_2$- directions, considering as $x$- и $y$- directions as well)

$$\begin{cases} \dot{\varepsilon}_1 + v_1 \varepsilon'_1 + b_1(\varepsilon_1, \varepsilon_2)\varepsilon_1 = 0 \\ \dot{\varepsilon}_2 + v_2 \varepsilon'_2 + b_2(\varepsilon_1, \varepsilon_2)\varepsilon_2 = 0 \end{cases}, \quad \dot{\varepsilon}_{1,2} \equiv \frac{\partial \varepsilon_{1,2}}{\partial t}, \; \varepsilon'_1 = \frac{\partial \varepsilon_1}{\partial x}, \; \varepsilon'_2 = \frac{\partial \varepsilon_2}{\partial y}, \qquad (1)$$

where $v_{1,2}$ represents the modules of axial (quasi-axial) elastic wave's velocities, $b_{1,2}$ – the coefficients, that take into account «effective» wave attenuations in the $x$- и $y$- directions correspondently. In particular, the attenuation's compensation takes place only in the field of energy output, that displays by the dependences of $b_{1,2}$ from $\varepsilon_{1,2}$:

$$b_{1,2}(\varepsilon_1, \varepsilon_2) = æ_{1,2}\big(1 - \Theta(\varepsilon_1 - \varepsilon_{1th})\Theta(\varepsilon_{2th} - \varepsilon_2)\big). \qquad (2)$$

In (2) parameters $æ_{1,2} > 0$ represent wave's attenuations in the absence of enhancement mechanisms, $\Theta(\varepsilon)$ – Heaviside's function. Considering the threshold deformations an elastic description of the medium is acceptable, harmonic approximation is in use:

$$\varepsilon_1 = (\varepsilon_1)_{max} k_1 \cos(\omega_1 t - k_1 x), \; \varepsilon_2 = -(\varepsilon_2)_{max} k_2 \cos(\omega_2 t - k_2 y), \qquad (3)$$

where $(\varepsilon_{1,2})_{max}$ – amplitudes, $\omega_{1,2}$ – frequencies of vibrations, connected with wave numbers $k_{1,2}$ by the standard relation $\omega_{1,2} = v_{1,2} k_{1,2}$. Picture 1 illustrates the scheme of the crystal growth with initially exited cell of the extended rectangular parallelepiped form with axial dimensions $d_{1,2} < \lambda_{1,2}/2$ ($\lambda_{1,2}$ – corresponding waves lengths); with $t = 0$ deformations are extreme in the cells' center ($x = y = 0$) and $\varepsilon_{1,2} = 0$ by $x = \pm\frac{\lambda_1}{4}, y = \pm\frac{\lambda_2}{4}$.

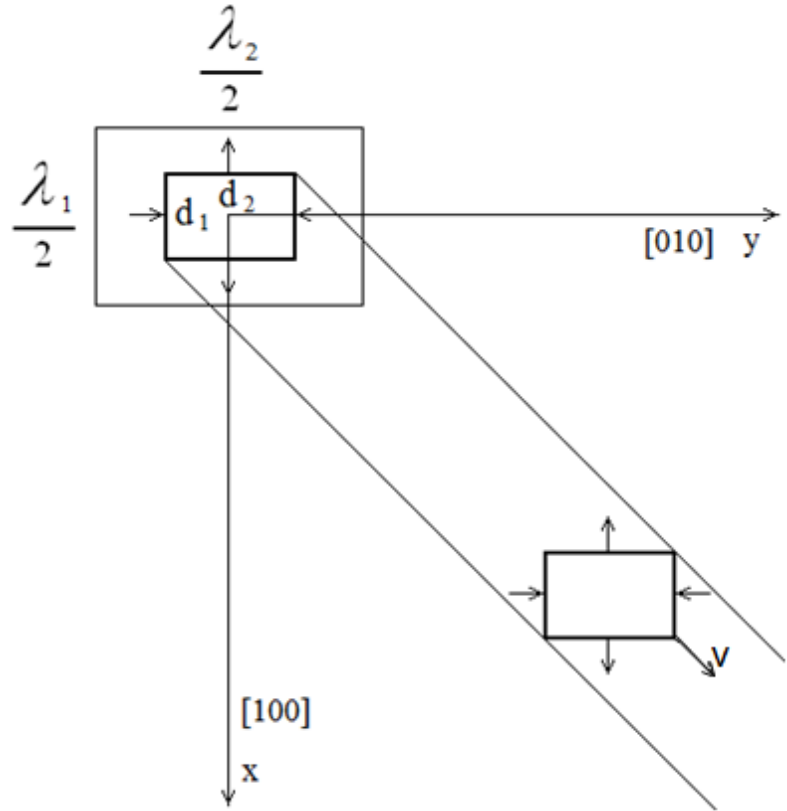

Pic. 1. Forming of the plate-like-type field, that loses its stableness with propagation of the threshold deformation.

Let $d_{1,2}$ be equal to $\lambda_{1,2}/4$. In this case, on the edges of the effective cell the deformations are lower approximately for 29% in comparison with the central point. The inequalities $\varepsilon_1 > \varepsilon_{1\text{th}}$, $|\varepsilon_2| > |\varepsilon_{2\text{th}}|$ take place in the inner side of the cell, whereas initial conditions of the deformation are of the form:

$$(\varepsilon_1)_{\max} \cos(k_1 x) \geq \varepsilon_{1\text{th}}, \quad -\frac{\lambda_1}{8} \leq x \leq \frac{\lambda_1}{8}; \quad |\varepsilon_2|_{\max} \cos(k_2 y) \geq |\varepsilon_{2\text{th}}|, \quad -\frac{\lambda_2}{8} \leq y \leq \frac{\lambda_2}{8}. \tag{4}$$

Considering that in the area, where overlay the propagation and compression deformations, $b_{1,2} = 0$ and taking into account the initial conditions (4), we obtain from (3)

$$\varepsilon_1(\zeta_1,\zeta_2) = (\varepsilon_1)_{\max} \cos(k_1 \zeta_1)\, \varphi(\zeta_1,\zeta_2), \quad \varepsilon_2(\zeta_1,\zeta_2) = (\varepsilon_2)_{\max} \cos(k_2 \zeta_2)\, \varphi(\zeta_1,\zeta_2), \tag{5}$$

where $\zeta_1 = x - v_1 t$, $\zeta_2 = y - v_2 t$, and function $\varphi(\zeta_1,\zeta_2) = [\Theta(\zeta_1 - d_1/2) - \Theta(\zeta_1 + d_1/2)][\Theta(\zeta_2 - d_2/2) - \Theta(\zeta_2 + d_2/2)]$ describes the motion of the exited condition with velocity $\mathbf{v} = \mathbf{v}_1 + \mathbf{v}_2$, according to the Pic. 1. Now it is quite easy to write down the generalization for the effective attenuation (2):

$$b_{1,2}(\varepsilon_1, \varepsilon_2) = æ_{1,2}\left(1 - \frac{\sigma_0}{\sigma_{th}} \Theta(\varepsilon_1 - \varepsilon_{1\text{th}}) \Theta(\varepsilon_{2\text{th}} - \varepsilon_2)\right), \tag{6}$$

where $\sigma_0$ и $\sigma_{th}$ represent the initial and threshold inverse occupancy (ref. [2]). By $\sigma_0 > \sigma_{th}$ the wave-generation process takes place and since then we have $0 < b_i = æ_i\left(1 - \frac{\sigma_0}{\sigma_{th}}\right)$ and CWP can be characterized by the deformations, that exceed the initial values.

While making the description of the crystals' thickness (profile) changes, it can be essential to consider, that the approximation to the space heterogeneity during the CWP-propagation can be followed by the increase of the $\sigma_{th}$ at the expense of the intensification of the attenuation for phonons and electrons and the reduction of the quantity of electrons, that take part in the generation process. This fact leads us for some extra coordinate-depended contribution to the effective attenuation (6) oh the controlling waves. Thus it is natural to expect the changes of the $d_{1,2}$ dimensions for the active cell: they are defined by the CWP-boarders. Thereby, while lengths of the waves $\lambda_{1,2}$ are inalterable in composition of CWP, the execution of the inequalities $d_1 < \lambda_1/2$, $d_2 < \lambda_2/2$ provides the reduction of the $d_{1,2}$ in case of the growth of the waves attenuation (in the course of approach to the heterogeneity), and the enhancement of the $d_{1,2}$ in case of the propagation of the waves attenuation (in the course of moving away from the heterogeneity). For some illustrative purposes while making the analytic description of the values' $d_{1,2}$ changes, let us consider, $d_1 = d_2$, $|\mathbf{v}_1| = |\mathbf{v}_2|$, $æ_1 = æ_2 = æ$, centers of heterogeneity localizations coincide with the planes $x = \bar{x}$, $y = \bar{y}$, $|\mathbf{v}| = \sqrt{2}|\mathbf{v}_{1,2}|$. Dependences of the $b_i$ from the coordinates are analogical:

$$b_1(\varepsilon_1, \varepsilon_2, x) = æ\left[1 - \frac{\sigma_0}{\sigma_{th}}\Theta(\varepsilon_1 - \varepsilon_{1th})\Theta(\varepsilon_{2th} - \varepsilon_2) + \delta_0 f\left(\frac{x-\bar{x}}{\lambda}\right)\right],$$
$$b_2(\varepsilon_1, \varepsilon_2, y) = æ\left[1 - \frac{\sigma_0}{\sigma_{th}}\Theta(\varepsilon_1 - \varepsilon_{1th})\Theta(\varepsilon_{2th} - \varepsilon_2) + \delta_0 f\left(\frac{y-\bar{y}}{\lambda}\right)\right]. \tag{7}$$

«Intensities» of the heterogeneities can be characterized with the help of the parameter $\delta_0$, and the introduction of the wave $\lambda$ to the measuring of the distances in the units of $\lambda$. To perform the further analysis it is sufficient to investigate one equation from (3), considering in (7) $\Theta(\varepsilon_1 - \varepsilon_{1th})\Theta(\varepsilon_{2th} - \varepsilon_2) = 1$, $\sigma_{th} - \sigma_0 \leq 0$ and omitting indexes 1, 2 for deformations in the orthogonal directions, that is

$$\dot{\varepsilon} + v\varepsilon' + b(x)\varepsilon = 0, \quad b(x) = æ\left[1 - \frac{\sigma_0}{\sigma_{th}} + \delta_0 f\left(\frac{x-\bar{x}}{\lambda}\right)\right]. \tag{8}$$

An obtained with the help of (8) value $d_{1,2}$ is $\sqrt{2}$ times less than the thickness of the crystal prototype $d$. All over further in the pictures the dependence $d_1(x)$ will be considered. It is natural to look for the decision (8) of the type of $\varepsilon(x,t) = \tilde{\varepsilon}(x,t)\cos(\omega t \pm kx)$ considering the amplitudes $\tilde{\varepsilon}(x,t)$ as sufficiently slow in variations. In this case in the first approximation it becomes possible to neglect the frequency $\omega$ changes, while signs $\pm$ in recording of the wave phase represent the possibility of the propagation towards $\pm \mathbf{k}$ directions. Considering the condition $v = \omega k$, from (8) we then obtain the equation for the amplitudes $\tilde{\varepsilon}(x,t)$:

$$\dot{\tilde{\varepsilon}} + v\tilde{\varepsilon}' + b(x)\tilde{\varepsilon} = 0, \tag{9}$$

where for some definiteness $+\mathbf{k}$ was picked as the direction of the propagation. Considering that at the initial time moment exited state area includes the field with the point $x = 0$, let us assume $\tilde{\varepsilon}(0,0) = \tilde{\varepsilon}_0$ and investigate the following versions of the $f$-function in (8):

$$f\left(\frac{x-\bar{x}}{\lambda}, \delta\right): \text{ a) } \exp\left(\frac{-|x-\bar{x}|}{\lambda}\right), \text{ b) } \left(\frac{(x-\bar{x})^2}{\lambda}\right), \text{ c) } \left(\frac{(x-\bar{x})^{-2}}{\lambda}\right). \tag{10}$$

For the estimation of the dimension $d(x)$, it is sufficient to find the static solution of the equation (9). Choosing e.g. the exponential dependence of the $f$-function in (10), from (9) with $\dot{\tilde{\varepsilon}} = 0$ we get

$$\begin{cases} \tilde{\varepsilon} = \tilde{\varepsilon}_0 \exp\left[-\frac{æ}{v}\left(1 - \frac{\sigma_0}{\sigma_{th}}\right)x - \frac{\lambda æ \delta_0}{v}\exp\left(-\frac{\bar{x}}{\lambda}\right)\left[\exp\left(\frac{x}{\lambda}\right) - 1\right]\right], & x \leq \bar{x} \\ \tilde{\varepsilon} = \tilde{\varepsilon}(\bar{x}) \exp\left[-\frac{æ}{v}\left(1 - \frac{\sigma_0}{\sigma_{th}}\right)(x - \bar{x}) - \frac{\lambda æ \delta_0}{v}\left[1 - \exp\left(\frac{-(x-\bar{x})}{\lambda}\right)\right]\right], & x > \bar{x} \end{cases} \tag{11}$$

Here $\tilde{\varepsilon}(\bar{x})$ we obtain with the substitution of $\bar{x}$ in (11) in case $x \leq \bar{x}$:

$$\tilde{\varepsilon}(\bar{x}) = \tilde{\varepsilon}_0 \exp\left[-\frac{æ}{v}\left(1 - \frac{\sigma_0}{\sigma_{th}}\right)\bar{x} - \frac{\lambda æ \delta_0}{v\delta}\left[1 - \exp\left(-\frac{\bar{x}}{\lambda}\right)\right]\right]. \tag{12}$$

Considering, according to (4), that the board of the controlling process in the $x$-direction is specified by the condition $\varepsilon_{th} = \tilde{\varepsilon}(x)\cos(kd(x)/2)$ from (11) by $k = 2\pi/\lambda$, we obtain:

$$\frac{d(x)}{\lambda} = \frac{1}{\pi}\arccos(\varepsilon_{th}/\tilde{\varepsilon}(x)) =$$

$$= \frac{1}{\pi}\begin{cases} \arccos\left\{\frac{\varepsilon_{th}}{\tilde{\varepsilon}_0}\exp\left[-\frac{\ae}{v}\left(\frac{\sigma_0}{\sigma_{th}}-1\right)x + \frac{\lambda\ae\delta_0}{v}\exp\left(-\frac{\bar{x}}{\lambda}\right)\left[\exp\left(\frac{x}{\lambda}\right)-1\right]\right]\right\}, & x \le \bar{x} \\ \arccos\left\{\frac{\varepsilon_{th}}{\tilde{\varepsilon}(\bar{x})}\exp\left[-\frac{\ae}{v}\left(\frac{\sigma_0}{\sigma_{th}}-1\right)(x-\bar{x}) + \frac{\lambda\ae\delta_0}{v}\left[1-\exp\left(\frac{-(x-\bar{x})}{\lambda}\right)\right]\right]\right\}, & x > \bar{x} \end{cases} \quad (13)$$

The parameter $\frac{\lambda\ae}{v}$ considered to be dimensionless, so with $\ae \approx 10^{-3}\frac{\omega}{2\pi}$ we have $\frac{\lambda\ae}{v} \approx 10^{-3}$. For the chosen initial state we assume $\frac{\varepsilon_{th}}{\tilde{\varepsilon}_0} = \frac{\sqrt{2}}{2}$. According to [2], let us assume $\frac{\sigma_0}{\sigma_{th}} - 1 \approx 1$, then the expression $\frac{\ae}{v}\left(\frac{\sigma_0}{\sigma_{th}}-1\right)x \approx \frac{10^{-3}x}{\lambda}$. Substituting $\tilde{x} = \frac{x}{\lambda}$ and $\tilde{d}_1(\tilde{x}) = \frac{d_1(x)}{\lambda}$ into (13), we finally get:

$$\tilde{d}_1(\tilde{x}) = \frac{1}{\pi}\begin{cases} \arccos\left\{\frac{\sqrt{2}}{2}\exp\left[-10^{-3}\tilde{x} + 10^{-3}\delta_0\exp(-\tilde{\bar{x}})\left[\exp(\tilde{x})-1\right]\right]\right\}, & x \le \bar{x} \\ \arccos\left\{\frac{\sqrt{2}}{2\tilde{\varepsilon}(\bar{x})}\exp\left[-10^{-3}(\tilde{x}-\tilde{\bar{x}}) + 10^{-3}\delta_0\left[1-\exp(-(\tilde{x}-\tilde{\bar{x}}))\right]\right]\right\}, & x > \bar{x} \end{cases} \quad (14)$$

Picture 2 illustrates an example of the essential variation of the $\tilde{d}_1(\tilde{x})$. Crystal contraction (Pic. 2a) is specified by the fields' origin ($\tilde{\bar{x}} = 10$), where the maximum attenuation takes place. Pic. 2b demonstrates, that even the negligible increment (in comparison with the previous case (a)) of the parameter $\delta_0$ brings us to the state of qualitative change of the solution: the right side Pic. 2b represent either the case of the initial state of excitement, which center $x$ can be localized in any point of the field $x^* \le x < \infty$ (for $x \to \infty$, $d \to \lambda/2$) in case the crystal grow towards the reduction of $x$, or the case of the initial state of excitement, which center $x$ can be localized in point $x^*$, in case the crystal grow towards the enhancement of $x$; the left side Pic. 2b describe the typical behavior of the $\tilde{d}_1$ in the immediate vicinity to the center of the heterogeneity.

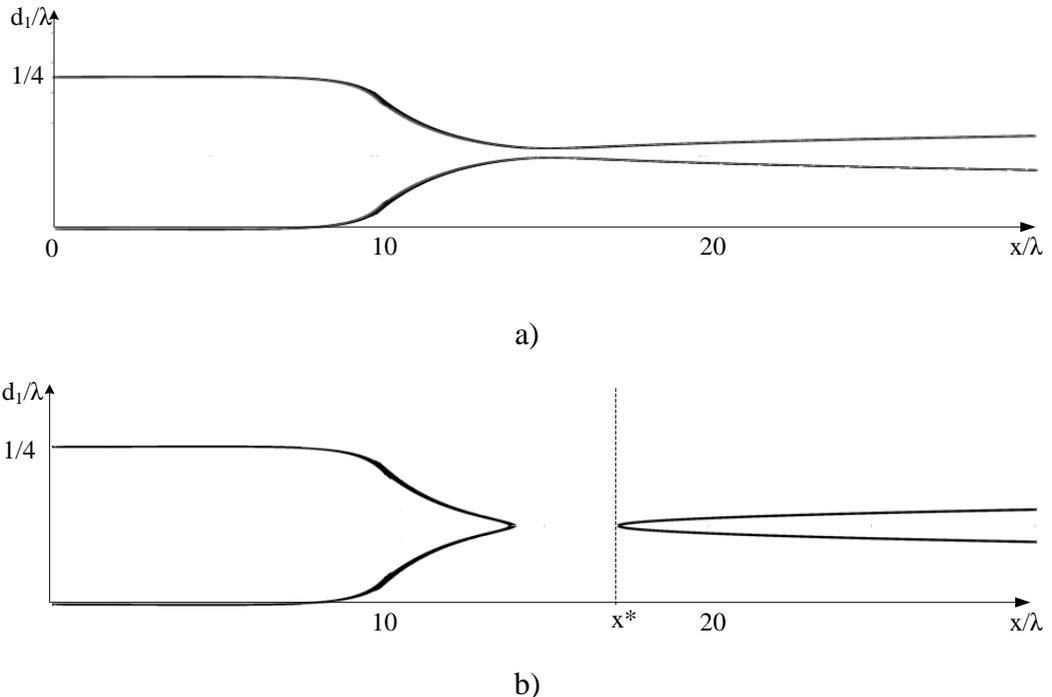

Pic. 2. Dependence $\tilde{d}_1$ considering with the parameters $\tilde{\bar{x}} = 10$, $\frac{\ae}{v}\left(\frac{\sigma_0}{\sigma_{th}}-1\right) = 10^{-3}$. a) $\delta_0 = 1{,}81 \cdot 10^2$, b) $\delta_0 = 1{,}82 \cdot 10^2$.

Let us notice, that due to the ocular demonstration of the crystal shape changes on the big spatial interval along the growth direction, representing the long crystal axe, the scales of the orthogonal directions on the Pic. 2 are sufficiently different. Picture 2 illustrates the dependences $\tilde{d}_1(\tilde{x})$, that were obtained with the substitution into (9) of some quadratic and inverse-quadratic functions of the type $f\left(\frac{x-\bar{x}}{\lambda}\right)$ into (10).

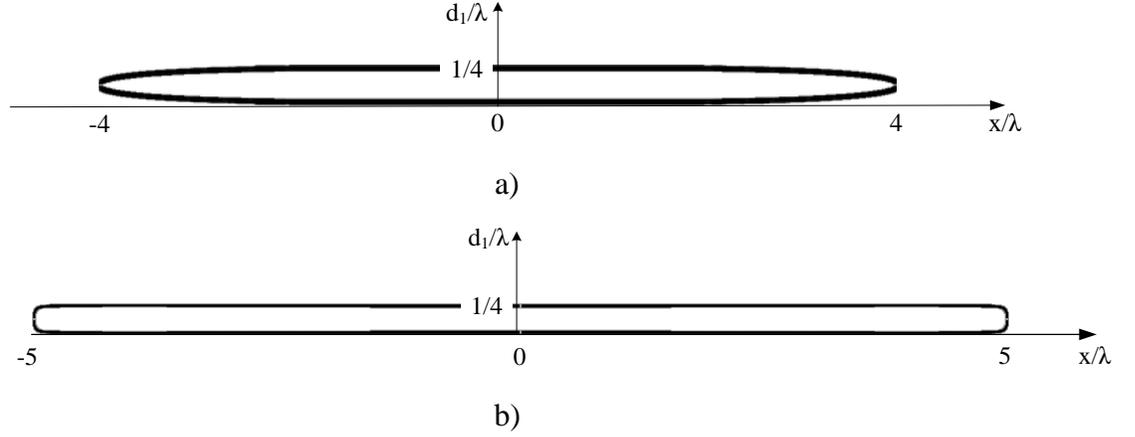

Pic. 3. Samples of dependences $\tilde{d}_1$: a) $f\left(\frac{x-\bar{x}}{\lambda}\right) = \left(\frac{(x-\bar{x})^2}{\lambda}\right)$, $\delta_0 = 10^2$, $\tilde{x} = 0$, b) $f\left(\frac{x-\bar{x}}{\lambda}\right) = \left(\frac{(x-\bar{x})^{-2}}{\lambda}\right)$, $\delta_0 = 10$, $\tilde{x} = 5$.

For the both modifications represented on the Pic. 3 the point $x = 0$ corresponds to the central point of the field of the initially exited state. The symmetry of the images in ratio to the replacement of $x$ to $-x$ means that under the consideration we now have the solutions, that response for the propagation of the wave beams from the initiation field towards to $\pm k$-directions. As soon as $f\left(\frac{x-\bar{x}}{\lambda}\right) = \left(\frac{(x-\bar{x})^{-2}}{\lambda}\right)$ and the plane $x = \bar{x}$ corresponds to the superior barrier, an essential difference of the diagrams' shapes in the field of the butts of the crystal growth (nearby the points $x/\lambda = \pm 4$, Pic. 3a and $x/\lambda = \pm 5$, Pic. 3b) takes place.

The conversion from the threshold to the finishing deformations (with the help of the shear) leads us to the representative set of the crystal profiles to perform the comparison with the experimental data. Thereby, the analysis of the crystal profiles can be used for the rearrangement of the nature of the space heterogeneity of the pattern.